\documentclass[
aps,
prb,
twocolumn,
superscriptaddress,%
showpacs,%
preprintnumbers,%
byrevtex,%
floatfix%
]{revtex4}

\usepackage{dcolumn}
\usepackage{bm}
\usepackage{ifthen}
\usepackage[usenames]{color}
\usepackage{amssymb}
\usepackage{textcomp}
\usepackage{amsfonts}
\usepackage{amsmath}
\usepackage[dvips]{graphicx}
\begin{document}

\date{\today}
\title{Influence of extended defects on the formation energy, the hyperfine structure, and the zero-field splitting
of NV centers in diamond}

\author{Wolfgang K\"orner}
\email{wolfgang.koerner@iwm.fraunhofer.de}\affiliation{Fraunhofer Institute for Mechanics of Materials IWM, W\"ohlerstr. 
11, 79108 Freiburg, Germany}

\author{Daniel F. Urban}
\affiliation{Fraunhofer Institute for Mechanics of Materials IWM, W\"ohlerstr. 11, 79108 Freiburg, Germany}

\author{Christian Els\"asser}
\affiliation{Fraunhofer Institute for Mechanics of Materials IWM,
W\"ohlerstr. 11, 79108 Freiburg, Germany}
\affiliation{Albert-Ludwigs-Universita\"t Freiburg, Freiburger Materialforschungszentrum (FMF), Stefan-Meier-Straße 21, 79104 Freiburg, Germany}

\begin{abstract}
We present a density functional theory analysis of nitrogen-vacancy (NV) centers in diamond which are
located in the vicinity of extended defects, namely intrinsic stacking faults (ISF), extrinsic stacking faults (ESF), and coherent twin boundaries (CTB) on \{111\} planes in diamond crystals. 
Several sites for NV centers close to the extended defects are energetically preferred with respect to the bulk crystal. This indicates that NV centers may be enriched at extended defects. 
We report the hyperfine structure (HFS) and zero-field splitting (ZFS) parameters of the NV centers at the extended defects which typically deviate by about 10\% but in some cases up to 90\% from their bulk values. 
Furthermore, we find that the influence of the extended defects on the NV centers is of short range: NV centers that are about three double layers (corresponding to $\sim6$ $\mathring{A}$ away from defect planes already show bulk-like behavior.

\end{abstract}


\maketitle
\section{Introduction}
The negatively charged nitrogen vacancy (NV$^-$) in diamond consists of a substitutional
nitrogen atom with a vacant nearest-neighbor carbon site and an additional electron. 
This point-defect complex has many promising applications, for example as an almost atomic
probe for high spatial resolution magnetometry \cite{ba08,ma08,ac09}
or an optically addressable solid-state qubit for quantum computing.\cite{jac09}

Our motivation to study NV centers near extended defects like stacking faults (SF) and grain boundaries (GB) is twofold. In general, the properties of point defects are altered near such extended defects in a characteristic manner, which make the point defect a sensitive local probe for extended defects. In a recent study of Ivady et al.,\cite{iva19} the analysis of the electronic (hyperfine) structure enabled the assignment of previously unidentified divancancy-like color centers in silicon carbide (SiC) which have an improved robustness against photoionization and show room temperature stability.
A second motivation is related to ensembles of NV centers acting together as sensitive magnetic probe arrays, for example in the context of detection of microstructural defects in electronic devices or mechanical components. In order to achieve a high magnetic sensitivity, the ensemble of NV$^-$ centers needs to be spatially localized and aligned, and the coherence time (T$_2$) needs to be long enough (in the range of $\mu s$). Various groups of experimentalists could achieve highly aligned (up to 99\%) ensembles of NV centers by means of chemical vapor deposition (CVD).\cite{ish17,ost19} Here, NV centers are localized within only a few nanometer thin layers. Thermal treatment of such layers can lead to conversion of further substitutional nitrogen atoms into NV$^-$ centers and increase T$_2$ by annealing lattice imperfections. However, thermal annealing may also reduce the alignment or even destroy NV centers under some conditions.\cite{ish17,ost19,oza18}

The CVD nucleation and growth processes for diamond create extended crystal defects including twin boundaries, stacking faults and dislocations which are frequently oriented along \{111\} planes.\cite{kab18}
Therefore, we consider a selection of such extended defects and their interplay with NV centers in this work. The presence of an extended defect may facilitate the alignment of the NV centers and also lead to an enrichment (segregation) of NV$^-$ at the extended defect. 
For our study we selected the intrinsic SF (ISF; stacking sequence ABC{\bf AB}ABC of \{111\} layers), the extrinsic SF (ESF; stacking sequence ABC{\bf ABAC}ABC), and the coherent twin boundary (CTB; stacking sequence ABC{\bf A}CBA). We have built supercell models in which we have placed the NV centers at all possible sites relative to the planar defect and evaluated their total energies to clarify whether the sites are preferred with respect to the bulk crystal.      
Furthermore, we evaluated the hyperfine structure (HFS) and the zero-field splitting (ZFS) parameters. These provide the possibility to identify sites of different point defects since these quantities are accurately measurable with sensitive techniques such as optically detected magnetic resonance (ODMR).\cite{jac09,aco10}

The manuscript is organized as follows: In section \ref{sec:supercell} 
the computational details and the supercell models are described. In sections
\ref{sec:hyperfine}, \ref{sec:zfs} and \ref{sec:comp} we briefly summarize important properties and the calculation of the tensors of hyperfine structure parameters 
$\boldsymbol{A}_{ij}^I$ and zero-field splitting parameters $\boldsymbol{D}_{ij}$. 
The results of the total energy calculations are presented in Sec.~\ref{sec:results:energy}. 
The results for the ZFS and HFS calculations are reported and discussed in sections
\ref{sec:results:zfs} and \ref{sec:results:hyperfine}, respectively. 
A summary in Sec.~\ref{sec:summary} concludes the paper.

\section{Theoretical approach}\label{sec:theory}

\subsection{Supercell models}\label{sec:supercell}

As a reference and starting system for the study of NV centers at \{111\} extended defects we 
constructed a hexagonal supercell of the diamond single crystal  
consisting of  $6\times 6\times 3$ hexagonal unit cells (6 carbon atoms each) with the c-axis pointing in [111]-direction.  
It thus contains $6^3 \cdot 3$ = 648 atoms.
Cubic diamond (Ramsdell notation\cite{ram47} 3C) in [111]-direction shows a layered structure with the sequence unit ABC of periodic stacks of carbon layers. There also exists a hexagonal diamond-like polytype (Ramsdell notation\cite{ram47} 2H) with the stacking sequence unit AB of carbon layers, which is named Lonsdaleite (see e.g. Ref.\,\onlinecite{own92} of Ownby et al. for the different polytypes of carbon). Lonsdaleite was taken as a second reference system because in the vicinity of extended defects the NV centers are sometimes located in a locally hexagonal atomic neighborhood.
Our  Lonsdaleite supercell consists of  $6\times 6\times 5$ hexagonal unit cells (4 carbon atoms each)
with a total of 720 atoms in the supercell. 

\begin{figure}[]
      \centerline{\includegraphics[width=\columnwidth]{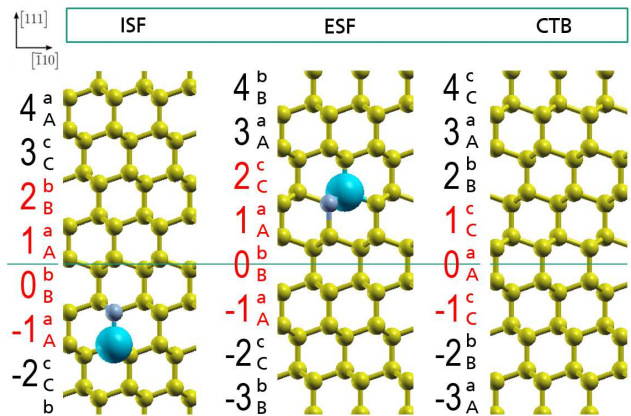}} 
      \caption{(Color online)
      Illustration of the atomic structure in the vicinity of the intrinsic stacking fault (ISF; one double layer Cc removed), the extrinsic stacking fault
      (ESF; one double layer Aa added) and the coherent twin boundary (CTB). For later reference we assign numbers to each double layer.
	  The yellow spheres represent the carbon atoms.       
      In the left panel an axial NV center in layer $n=-1$ at position aA is shown and the middle panel includes a basal NV center in position aC.  
      The position of the carbon vacancy is indicated by a large blue sphere and the position of the nitrogen atom is  
      represented by a small grey sphere. 
      The horizontal solid green line indicates the symmetry planes of the extended defects. 
      Note that the results displayed in the following figures will not be strictly symmetric because the NV defect complex 
      has a defined orientation and no symmetry center between the N atom and the C vacancy.
\label{fig:defect_models}}
\end{figure}

We have considered the three most prominent cases of extended planar \{111\} defects, namely the ISF, ESF, and CTB.
For the ISF we added two additional layers A and B to the bulk supercell (hence one C layer is missing). Since each double layer contains 72 atoms this supercell contains 648+144= 792 atoms in total. A close-up of the ISF region is shown in the left panel of Fig.~\ref{fig:defect_models}.
For the ESF, we inserted one additional layer A between a B and a C layer. This supercell contains 648+72= 720 atoms. The middle panel of 
Fig.~\ref{fig:defect_models} displays the atomic structure of the ESF region.
In order to obtain a CTB supercell which is periodic in c direction two properly misoriented crystalline layers are incorporated. Our supercell contains 16 double layers yielding 1152 atoms in total. The CTB region is shown in the right panel of Fig.~\ref{fig:defect_models}.

For the structural relaxation of the supercells mentioned above and
the calculation of hyperfine structure parameters we used the Vienna Ab Initio Simulation Package 
(VASP)\cite{kr96,kr99} version 5.4.4 with the projector-augmented-wave (PAW) method.\cite{bl94} 
The exchange-correlation interactions are described 
by the generalized gradient approximation (GGA) of Perdew, Burke and Ernzerhof (PBE).\cite{Perdew1996}       
The VASP calculations were carried out with a plane-wave cutoff energy of 420 eV for the valence electrons. 

For supercells with less than 800 atoms 
the Brillouin-zone integrals are evaluated on a $\Gamma$-centered $2\times 2\times 2$ k-point mesh with a Gaussian broadening of 0.05 eV.
For larger supercells we considered only the $\Gamma$-point. 
Atom positions were relaxed under constant volume condition until the residual forces acting on them were less than 0.01 eV/\AA\ and the energy difference
between two consecutive ionic relaxation steps was smaller than $10^{-5}$\,eV. 

We follow the work by Gali et al.\cite{gal09} and use a lattice constant of $a$= 3.567\AA\ for the primitive cubic cell of diamond which agrees well with
the experimentally determined one of Holloway et al..\cite{hol91} 
Consequently, the length of the two lattice vectors in the basal plane of the hexagonal structure is 15.13  \AA\ for all supercells.
The c-axis lattice vectors vary in length from 18.53 \AA\ (bulk diamond), 20.59 \AA\ (ESF and Lonsdaleite), 22.65 \AA\ (ISF) to 32.95 \AA\ (CTB).

In addition to the common ABC and AB notation of stacking sequences for close-packed cubic and hexagonal crystals, we also use the more detailed AaBbCc and AaBb notation for cubic and hexagonal diamond structures, in order to specify precisely the positions 
of the NV centers. The left panel of Fig-~\ref{fig:defect_models} includes a NV center with the nitrogen atom at an a-layer and the nearest-neighbor carbon vacancy at a layer A. NV centers that are oriented parallel to the [111] direction are denoted as \textit{axial}.  
The NV center in the middle panel of Fig.~\ref{fig:defect_models} is inclined by about 109.5 degree with respect to the c-axis and has the nitrogen atom in an a-layer and the carbon vacancy in a C-layer. NV centers in this orientation are denoted as \textit{basal}.

Note that the high C$_{3v}$ symmetry of the NV$^-$ is conserved for axially oriented NV centers near the extended defects.
However, the presence of a NV center in basal orientation reduces the symmetry to C$_{1h}$ which implies a further splitting of levels, as discussed in the following.

In order to determine the interplay of the extended defects and the individual NV centers we have 
placed the NV$^-$
defect at every symmetry-inequivalent axial and basal position within our supercells (with the N atom always in the layer a, b or c and the carbon vacancy in the layers A, B or C). For the discussion of results we assign numbers to each double layer and choose the number zero next to the symmetry plane of each extended defect (marked by the horizontal green line in Fig.~\ref{fig:defect_models}).

\subsection{Hyperfine interaction}\label{sec:hyperfine}
The hyperfine structure (HFS) tensor $\boldsymbol{A}^I$ describes the interaction between a nuclear spin S$_I$ 
(located at site $I$) and the electronic spin distribution. The hyperfine interaction between 
a nuclear spin S$_I$ (i.e., the nuclear spin of $^{13}$C, $^{14}$N or $^{15}$N)
and the electronic spin distribution S$_e$ (i.e., the spin of the defect state (q = -1) of the NV center) can be modeled with the Hamiltonian
\begin{equation}
H_{{\rm HFS}}= S_e  \boldsymbol{A}^I  S_I. 
\end{equation}
The hyperfine structure tensor components A$_{ij}^I$ for a nucleus $I$ are
\begin{equation}\label{eqn_HF}
A_{ij}^{I}=\frac{\mu_0 \gamma_{I} \gamma_e}{2S}\int d^3r ~ n_{S}({\bm r}) \left[ \left(\frac{8\pi \delta(r)}{3}\right) +\left( \frac{3x_ix_j}{r^5}-\frac{\delta_{ij}}{r^3}\right) \right],
\end{equation}
where the first term in brackets is the Fermi contact term and the second bracket is
the magnetic dipole-dipole term.
Here, $n_S$ denotes the spin density associated with spin state S,
$\mu_0$ is the vacuum magnetic permeability, $\gamma_e$ the gyromagnetic ratio of 
the electron and $\gamma_{I}$ the gyromagnetic ratio of the nucleus. In this work we use $\gamma{(^{13}{\rm C})}/2\pi$  = 10.7084 Mhz/T,
$\gamma{(^{14}{\rm N})}/2\pi$  = 3.077 Mhz/T and $\gamma{(^{15}{\rm N})}/2\pi$ = -4.316 Mhz/T, respectively.\cite{ber04}

The principal values $A_{11}$, $A_{22}$, and $A_{33}$ of the hyperfine tensor which can be found by
diagonalization are generally called hyperfine constants. If the hyperfine field
has C$_{3v}$  symmetry then $A_{11}=A_{22}=:A_{\perp}$ and $A_{33}=:A_{\parallel}$.

\subsection{Zero-field-splitting}\label{sec:zfs}
The ZFS arises as consequence of various interactions of the electronic states of an atom or ion in a molecule
or crystal due to the presence of more than one unpaired electron. The classic example for a
ZFS is that of a spin triplet for a S=1 spin system (as for the NV center in diamond).  
In the presence of an (externally applied) magnetic field, the degeneracy of the
levels with different values of magnetic spin quantum number (M$_S$= 0,+1,-1) is lifted (magnetic Zeeman splitting).
The effect of electron-electron repulsion by magnetic dipole-dipole interaction can
be modeled by the Hamiltonian
\begin{equation}\label{zfs}
H_{{\rm ZFS}}= \frac{\mu_0 g^2 \mu_B^2}{4\pi r^5}  \bigl[3( S_1\cdot r)(S_2\cdot r)-(S_1\cdot S_2)r^2\bigr],
\end{equation}
where $r=r_1-r_2$ is the spatial distance between the spins, $S_i =\frac{1}{2}[\sigma_x, \sigma_y, \sigma_z]$ is the spin operator
vector of particle i, $\sigma_j$ (j=x, y, z) are the Pauli matrices, and g is the Land\'e factor.
One can separate the spatial and spin dependencies in equation (\ref{zfs}) and write the Hamiltonian in the form 
\begin{equation}
H_{{\rm ZFS}}= S  \boldsymbol{D} S,
\end{equation}
where $S=S_1+S_2$ is the total spin and $\boldsymbol{D}$ describes the dipolar spin-spin interaction.

It can be shown that the tensor $\boldsymbol{D}$ is symmetric and traceless and thus can be diagonalized. 
In the following, we name the diagonal elements $D_{ii}$. 
In general, one can split $H_{{\rm ZFS}}$ into a longitudinal component of the magnetic dipole-dipole interaction 
(named $D$) and a transversal component (named $E$). The Hamiltonian then reads
\begin{equation}
H_{{\rm ZFS}}= D \bigl[S_z^2-\frac{1}{3}S(S+1)\bigr]+ E(S_x^2-S_y^2) 
\end{equation}
with 
\begin{equation}
D= \frac{3}{2} D_{zz} ~~ {\rm and} ~~  E=  \frac{1}{2}(D_{xx}-D_{yy}).
\end{equation}
The quantities $D$ and $E$ are accurately measurable with techniques such as ODMR\cite{jac09,aco10} or electron paramagnetic resonance (EPR).\cite{fel09,lou78}

An isolated NV center in a diamond crystal has C$_{3v}$ symmetry. This axial symmetry implies $D_{xx}$= $D_{yy}$
and thus $E = 0$.
When a NV center is close to an extended defect this degeneracy can be lifted significantly, as we will discuss later.
Of course, the same statement also holds for hyperfine structure constants and for axial symmetry $A_{xx}$= $A_{yy}$.

For more detailed information on the ZFS and HFS in diamond see e.g. Refs.\,\onlinecite{lou78} and \onlinecite{iva14}.

\subsection{Computational details concerning $\boldsymbol{A}_{ij}^I$ and $\boldsymbol{D}_{ij}$}\label{sec:comp}

The computation of the hyperfine structure tensor components $\boldsymbol{A}_{ij}^I$ and the zero-field splitting tensor components $\boldsymbol{D}_{ij}$ was done with routines implemented in VASP.
For the hyperfine structure tensor we follow the procedure of Sz\'asz et al.
who carried out a careful analysis of the role of core spin polarization on the hyperfine structure constants.\cite{sza13}
They found that the underestimation of the localization of the wavefunction by the GGA functional PBE in general leads to wrong results for the hyperfine structure constants A$_{ij}$.
Whereas using the hybrid functional HSE06\cite{hey03,kru06}, which yields more localized wavefunctions, leads to results which differ only by a few percent from the experimental values.
However, by neglecting the contribution of the core electrons (named A$_{1c}$) in the PBE calculations one can obtain a quite good agreement with the experimental values as well.
Due to the size of our supercell models ($\approx 1000$ atoms) we are restricted to the use of PBE and therefore apply the recipe of Sz\'asz et al..\cite{sza13} 
The quality of our results presented in Sec.~\ref{sec:results} demonstrates that this approximation is apparently sufficient. 
  
Well-relaxed structures are required as a prerequisite for calculating HFS constants as the HFS is sensitive to small structural changes.
For the NV defect complex in the bulk diamond supercell, we compared the results for A$_{ii}$ and D$_{ii}$ for structures with forces relaxed to less than 0.01 eV/\AA\ and 0.001 eV/\AA, respectively. The values of the diagonal elements D$_{ii}$ of the ZFS tensor differed only by about 0.1\% between the two cases. The values of the diagonal elements A$_{ii}$ of the HFS tensor for $^{14}$N differed by up to 1.2\% and for $^{13}$C by approximately 0.3\%.
The A$_{ii}$ for $^{14}$N have the largest variations since they have the smallest absolute values.
For the smaller supercells we also performed convergence tests for the ZFS and HFS constants with respect to k-mesh and plane-wave cutoff parameters. We verified that a $\Gamma$-point calculation and a plane-wave cutoff energy of 420 eV is already sufficient for numerically converged results for the supercells with the extended defects.

\section{Results and discussion}\label{sec:results}

All results reported in the following were obtained for negatively charged NV centers (NV$^-$).

\subsection{Formation energies of NV centers at extended defects}\label{sec:results:energy}
\begin{figure}[]
      \centerline{\includegraphics[width=\columnwidth]{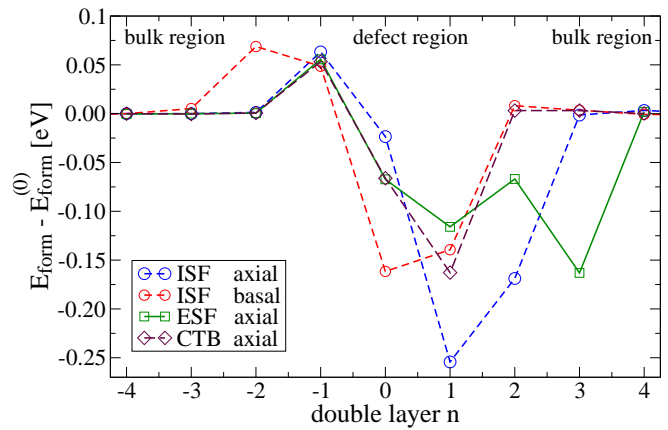}} 
      \caption{(Color online)
      Formation energies of NV centers with either axial or basal orientation in vicinity of an ISF, ESF, and CTB. 
      Values are given in eV and relative to the formation energy
      in bulk diamond, E$^{(0)}_{{\rm form}}$. 
\label{fig:form_energies}}
\end{figure}

Figure \ref{fig:form_energies} shows a selection of formation energies of NV centers with either axial or basal orientation 
at different positions relative to the considered extended planar defects. 
NV centers located at the double layers n=0, 1, 2 or 3 have lower energies than in the bulk crystal and therefore are energetically favorable. The minimum value of approximately $-0.25$ eV is obtained for a NV$^-$ in the axial position in double layer n= 1 at the ISF.
For the double layers n= -1 and n= -2 most of the formation energies have slightly positive values.
This asymmetry with respect to the defect plane originates in the orientation of the NV which is illustrated in Fig.~\ref{fig:defect_models}.

An interesting question concerning extended defects is whether they can be used as structural templates
for arranging an array of well aligned and equally oriented NV centers. This would require a significant difference between the formation energies
of the basal and axial orientation in addition to the condition that the formation energy of a NV center is lower at the extended defect than in the bulk crystal. 
For the ISF at the double layer n= 1 we found the largest difference in formation energies of approximately 0.1 eV. Since at room temperature
0.1 eV corresponds to a thermal energy of only 4k$_B$T we cannot expect a strict ordering of NV centers at ISF. For the ESF
and the CTB the differences of the axial and basal orientations are even smaller. (We refrained from plotting these results
for the basal orientation in the following figures for the sake of clarity.)

\subsection{Zero field splitting}\label{sec:results:zfs}

\subsubsection{Longitudinal ZFS component D} 
For a NV$^-$ center located in an otherwise perfect diamond crystal the splitting of the singlet and triplet states of the $^3A_2$ ground state was experimentally determined to high precision by Felton et. al.\cite{fel09}:  $D$=$\frac{3}{2}D_{zz}$= 2.872(2) GHz.
Our theoretical bulk value is $D_0$= 3.049 GHz and deviates by 6.2\% from the measured value.

\begin{figure}[]
      \centerline{\includegraphics[width=\columnwidth]{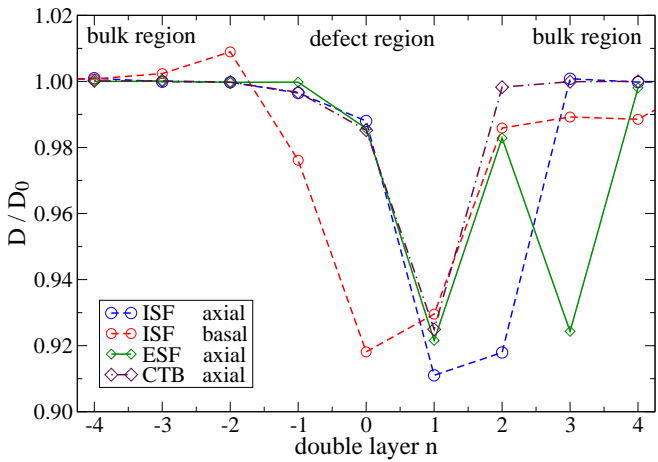}} 
      \caption{(Color online)
      Longitudinal component $D$ of the ZFS of a NV$^-$ center for different positions relative to the ISF, ESF and CTB. 
      Values are given with respect to the value $D_0$ of bulk diamond. 
\label{fig:Dzz_zfs}}
\end{figure}

In Fig.~\ref{fig:Dzz_zfs} the axial component $D$ of the ZFS normalized to the bulk
value $D_0$ is shown. At the ISF the axial oriented NV centers at the layers n = 1 and n = 2 have significantly reduced $D$ values. At first glance these results are astonishing as one would expect
this dip in $D$ to be symmetric with respect to the center of the ISF. However, the main contribution to the ZFS originates from the three
C atoms next to the vacancy, since they each have a magnetic moment of about 0.52$\mu_B$ (in summary corresponding to 78\% of 2$\mu_B$).
For the NV center located in the double layer n = 1 these carbon atoms are situated in layer b of the double layer n = 0 (cf. Fig.~\ref{fig:defect_models}).
Note that the double layers 0 and 1 of the ISF are locally in a hexagonal stacking ABA and BAB.   
The observed decrease of about 9\% is consistent with the results $D_0$= 2.681 GHz obtained for hexagonal Lonsdaleite (with AB stacking). The latter value corresponds to about 88\% of the value for diamond. 

All the other dips observed in Fig.~\ref{fig:Dzz_zfs} can be explained in the same way by NV centers where
the three carbon atoms next to the vacancy lie in a locally hexagonal layer-stacking sequence. At the ESF the two ABA stacking sequences are separated by one double layer, leading to a W-shape of the data, while the curve for the CTB shows a V-shape since there only is a single ABA stacking sequence.
 
\subsubsection{Transversal ZFS component E}
\begin{figure}[]
      \centerline{\includegraphics[width=\columnwidth]{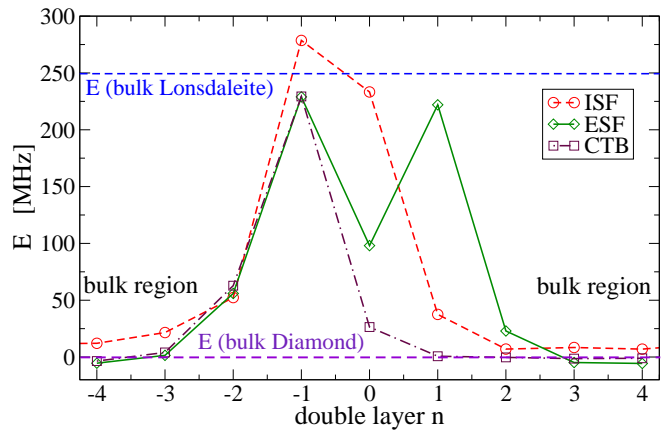}} 
      \caption{(Color online)
Transversal component E of the ZFS (in MHz) for a NV-center with basal orientation at different positions relative to the ISF, the ESF and the CTB.
\label{fig:E_zfs}}
\end{figure}

Figure \ref{fig:E_zfs} shows the calculated values of the transversal component $E$ of the ZFS. The basal orientation of the NV center in combination with the extended defect breaks the 
threefold symmetry and thus $E$ becomes nonzero. The influence of the extended defects on the transversal component $E$ is apparently of short range and the $E$ values quickly decay to zero with growing distance to the planar defect. 
\footnote{The value zero is not exactly obtained due to the finite size of the supercells with periodic boundary conditions. 
A basal NV center in our  $6\times 6\times 3$ hexagonal supercell of the diamond bulk crystal has a finite value of $E_0=$ 25 MHz. Ideally $E_0$ should be zero in this case. However, the value zero is only numerically obtained for axially oriented NV-centers in a hexagonal supercell for which the c axis is aligned with the NV-center and the full symmetry is conserved. The values given in Fig.~\ref{fig:E_zfs} are corrected by this offset $E_0$ (which also explains the slightly negative $E$ values for the ESF and the CTB in the figure).} 

Significant deviations in $E$ from its bulk value occur whenever the three C atoms next to the C vacancy are located in an ABA stacked neighborhood.
For the ISF this is the case e.g. for the NV center at the double layer n= -1, where $E$ is about 278 MHz, and at n= 0, with 233 MHz which is close to the value obtained for the hexagonal Lonsdaleite, namely 249 MHz. It is conceivable that basally oriented NV centers at double layers n= -1 and n= 0 show approximately the transversal ZFS of Lonsdaleite. For the double layer n= -1 two carbon atoms are in the $a$-layer of double layer n= -1 and one carbon atom is in the $b$-layer of double layer n= 0. This again results in a local hexagonally stacked neighborhood ABA (cf. Fig.~\ref{fig:defect_models}). For the double layer n= 0 one carbon atom is in the $a$-layer of double layer n=1 while two carbon atoms are in the $b$-layer of double layer n= 0, which yields the local hexagonally stacked neighborhood BAB.

The reasoning is similar for basal NV centers at the ESF and the CTB which explains the (inverted) W- and V-shapes of the data sets. Like in the case of the longitudinal component $D$ the magnitude of the extremal values for ESF and CTB are very similar and can be explained by the appearance of local hexagonal ABA or BAB stackings.

The finite $E$ value leads to a splitting of the triplet. Including an external magnetic field $B_{{\rm ext}}$ the energy levels for a spin S=1 system are:\cite{boc04}
$E_{{\rm triplet}}=D\pm\sqrt{E^2+(g \mu_B B_{{\rm ext}})^2}$ and $E_{{\rm singlet}}=0$.

\subsection{Hyperfine structure}\label{sec:results:hyperfine}

\subsubsection{Hyperfine splitting connected to $^{14}{\rm N}$ and $^{15}{\rm N}$ } 
Before discussing the results for the hyperfine structure constants at the three considered
extended defects we compare our results for the NV center in the cubic diamond crystal to other
theoretical and experimental results. In table \ref{N_A_Bulk} the hyperfine structure parameters
for $^{14}$N and $^{15}$N are given. 
In the more recent experimental work\cite{fel09}, Felton et al. determined the sign of the parameters by measuring electron-nuclear
double spin-flip transitions and apparently provide more accurate hyperfine parameters than He et al.\cite{he93}
For more details we refer the reader to the discussion given in Ref.\,\onlinecite{fel09}.

For all hyperfine structure constants the different theoretical studies agree with the experimental work of Felton et al.\cite{fel09} regarding the signs but not in the absolute values for which there are some considerable deviations. 
Gali et al.\cite{ga08} report an isotropic hyperfine interaction for $^{14}$N since their $b=(A_{\parallel}-A_{\perp})/3$ is zero. By contrast, our value is $b\approx 0.14$ MHz which is in the range of 0.19(7) MHz measured by Felton et al.
For $^{15}$N, both Gali et al.\cite{ga08} and we find an anisotropic interaction. Ref.~\onlinecite{ga08} reports $b\approx -0.13$ MHz, while our value is $b\approx -0.20$ MHz which again is very close to the measured value $-0.21$(3) MHz of Ref.~\onlinecite{fel09}.

The work of Sz\'asz et al.\cite{sza13} basically agrees with our absolute values and the little deviations stem
from the use of different supercell models. Our results are obtained from the 648-atom hexagonal supercell whereas they have used a 512-atom simple cubic supercell.

The theoretical results obtained with the HSE06 functional on the 512-atom supercell show the least deviations from the absolute experimental values of Felton et al. but the ratio $A_{\perp}$/$A_{\parallel}$ deviates.    
Our calculated ratios $A_{\perp}$/$A_{\parallel}$ for $^{14}$N and $^{15}$N  agree reasonably well with the experimental data.

\begin{table}[]
\begin{tabular}{l c c r }
\hline \hline 
Defect in bulk diamond ~& ~~ $A_{\parallel}$ [Mhz]~~ &~~ $A_{\perp}$ [Mhz] ~ & $A_{\perp}$/$A_{\parallel}$ \\
\hline 
$^{14}$NV$^{-}$ (this work, PBE)    & -1.73   & -2.16     & 1.25   \\
$^{14}$NV$^{-}$ (exp.)$^{a}$        & -2.14(7)& -2.70(7)  & 1.26   \\
$^{14}$NV$^{-}$ (exp.)$^{b}$        & +2.30(2)& -2.10(10) & 1.10   \\
$^{14}$NV$^{-}$ (theo., LSDA)$^{c}$ & -1.7    & -1.7      & 1.00   \\
$^{14}$NV$^{-}$ (theo., PBE)$^{d}$  & -1.71   & -2.12     & 1.24   \\
$^{14}$NV$^{-}$ (theo., HS06)$^{d}$ & -2.38   & -2.79     & 1.17   \\
\hline
$^{15}$NV$^{-}$ (this work, PBE)    & 2.43    & 3.03      & 1.25   \\
$^{15}$NV$^{-}$ (exp.)$^{a}$        & 3.03(3) & 3.65(3)   & 1.21   \\
$^{15}$NV$^{-}$ (theo., PBE)$^{e}$  & 2.3     & 2.7       & 1.17   \\
\hline \hline
\end{tabular}
\caption{Comparison of hyperfine structure parameters $A_{\parallel}$ and $A_{\perp}$ for $^{14}$N and $^{15}$N in bulk diamond for the $^3A_2$ ground state. The theoretical and experimental data is taken from  $^{a}$Ref.\,\onlinecite{fel09}, $^{b}$Ref.\,\onlinecite{he93}, $^{c}$Ref.\,\onlinecite{ga08}, $^{d}$Ref.\,\onlinecite{sza13}
 and $^{e}$Ref.\,\onlinecite{ga09}. Our absolute values differ from the experimental values of Felton et al.\cite{fel09} but the
ratios $A_{\parallel}$/$A_{\perp}$ agree quite well.
\label{N_A_Bulk}}
\end{table}

\begin{figure}[]
      \centerline{\includegraphics[width=\columnwidth]{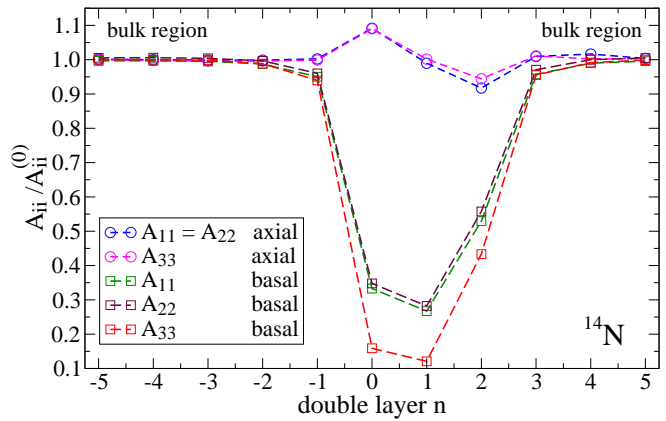}} 
      \caption{(Color online)
      Hyperfine structure constants $A_{ii}$ for $^{14}$N for the different positions at the ISF with either axial or basal orientation of the NV center. The $A_{ii}$ values are normalized by the corresponding values $A^{(0)} _{ii}$ of bulk diamond. 
\label{fig:A_n14}}
\end{figure}

Fig.~\ref{fig:A_n14} shows the results for the hyperfine structure constants for the ISF.
For axially oriented NV centers all three diagonal elements $A_{ii}$ (i=1,2,3) increase by about 10\% at layer n= 0
whereas they decrease by about 10\% at layer n= 2. However, the ratio $A_{\perp}/A_{\parallel}$ remains almost constant.
The situation is rather different for the basal orientation of the NV-center. Here, we observe a huge decrease of about +1.5 MHz for all three $A_{ii}$. 
For $A_{33}$ this corresponds to a reduction of 90\% in close vicinity to the defect plane.
The diagonal elements $A_{11}$ and $A_{22}$ decrease by about 70\% at layer n= 1 and show a small splitting of 0.02 to 0.06 MHz due to the presence of the ISF.

Of course, in the case of the nitrogen hyperfine interaction the location of the N atom (and not the location of the C atoms) is decisive. The strongest decrease of the hyperfine parameters occurs when the N atom is located in the double layer n= 0 or n= 1 where it is embedded in an ABA stacked neighborhood.

The $A_{ii}$-values for the ESF and CTB (not included in the figure) show the same magnitude of absolute variation as obtained for the ISF. The data show the characteristic W-shape for the ESF and V-shape for the CTB, as previously observed for the $D$ and $E$ components of the ZFS (cf. Figs.~\ref{fig:Dzz_zfs} and \ref{fig:E_zfs}, respectively).

\subsubsection{Hyperfine splitting connected to $^{13}{\rm C}$ } 

A compilation of hyperfine structure parameters
for $^{13}$C for bulk cubic diamond is listed in Tab.~\ref{tab_C13_A_Bulk} including our results and values taken from literature.
The comparison of the most recent theoretical PBE results of Gali et al.\cite{ga09}, Sz\'asz et al.\cite{sza13} and us to the most recent experimental results of Felton et al.\cite{fel09} yields a satisfactory agreement.
Interestingly the theoretical results obtained with the more sophisticated HSE06 functional deviate more from the experimental results than those obtained by using the PBE functional.
We therefore conclude that the latter provides an excellent starting point for studying the effects of the extended defects on the hyperfine splitting.
Note that in the early work of Loubser et al.\cite{lou78} electron paramagnetic resonance was used which does not allow to determine the sign of the hyperfine structure constants.

\begin{figure}[]
      \centerline{\includegraphics[width=\columnwidth]{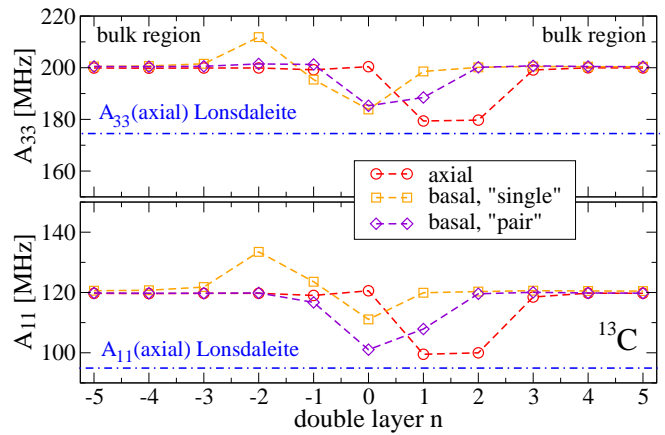}} 
      \caption{(Color online)
Hyperfine structure constants $A_{11}$ (lower panel) and $A_{33}$ (upper panel) for $^{13}$C for the various positions relative to the ISF with either axial or basal orientation of the NV center. For the basally oriented NV centers two carbon atoms lie in a common layer. Their values are denoted by "pair". The third carbon atom residing in the layer above the vacancy is denoted by "single".
\label{fig:A_c13}}
\end{figure}

As in the case of nitrogen we present the hyperfine interaction results only for the ISF in Fig.~\ref{fig:A_c13} 
since these are sufficiently representative.
For the axially oriented NV$^-$ center the strongest decrease of
the $A_{ii}$ is of about 20 MHz at 
double layers n= 1 and n= 2. Again, like in the case of the ZFS, the minimum is shifted by one double
layer to the right relative to the center of the ISF. This is conceivable since we look at the hyperfine interaction
of $^{13}$C atoms located at one of the three sites next to the C vacancy. Thus for an axial NV center situated at double layer n= 1
the respective $^{13}$C atom is situated in layer b of double layer n= 0 and locally in a hexagonal ABA neighborhood. The reduced $A_{ii}$ values are consistent with the ones obtained for Lonsdaleite (see blue dashed-dotted lines in Fig.~\ref{fig:A_c13}).

The behavior of the NV center with basal orientation is more complex. Like for $^{14}$N some $A_{ii}$ values increase and others
decrease. However, the $^{13}$C atom can be located in two different layers. There are two sites for carbon atoms in the same layer of the nitrogen atom, which we call "pair", and one site in the layer above the vacancy, which we denote by "single". The presence of an extended defect changes the charge distribution. Wherever the $A_{ii}$ values are lowered the underlying reason is a reduced charge density (see equation \ref{eqn_HF}). When $A_{ii}$ is higher than in the bulk we found an increased charge density at the respective $^{13}$C atom.    

Furthermore, we obtained a splitting of $A_{11}$ and $A_{22}$ by 0.1 to 0.2 MHz at the "pair" sites and by 
0.2 to 0.6 MHz at the "single" sites. 
This splitting originates in the breaking of the C$_{3v}$ symmetry for the basally oriented NV centers induced by the presence of the extended defects.
A similar splitting is observed when modeling the NV-center in bulk diamond in a cubic supercell, which also breaks the C$_{3v}$ symmetry (see supplemental material of Ref.~\onlinecite{sza13}).
Note the much smaller order of magnitude of this induced splitting compared to the absolute values of the $A_{ii}$. We did not include the $A_{22}$ components in Fig.~\ref{fig:A_c13} for the sake of clarity.

\begin{table}[t]
\begin{tabular}{l c c r }
\hline \hline 
bulk $^{14}$NV$^{-}$ center ~& ~~ $A_{\parallel}$ [Mhz]~~ &~~ $A_{\perp}$ [Mhz] ~ & $A_{\parallel}/A_{\perp}$ \\
\hline 
$^{13}$C (this work, PBE)    & 199.9    & 119.2     & 1.68   \\
$^{13}$C (exp.)$^{a}$        & 199.7(2) & 120.3(2)  & 1.66   \\
$^{13}$C (exp.)$^{b}$        & $\pm$205 & $\pm$123  & 1.67   \\
$^{13}$C (theo., LSDA)$^{c}$ & 185.4    & 109.9     & 1.69   \\
$^{13}$C (theo., PBE)$^{d}$  & 201.1    & 120.1     & 1.67   \\
$^{13}$C (theo., PBE)$^{e}$  & 199.8    & 119.8     & 1.67   \\
$^{13}$C (theo., HSE06)$^{e}$& 228.8    & 144.5     & 1.58   \\
\hline \hline
\end{tabular}
\caption{Comparison of hyperfine structure parameters for $^{13}$C in the cubic diamond crystal, where the $^{13}$C is located at one of the three carbon sites next to the vacancy.  The theoretical and experimental data is taken from the $^{a}$Ref.\,\onlinecite{fel09}, $^{b}$Ref.\,\onlinecite{lou78}, $^{c}$Ref.\,\onlinecite{ga08}, $^{d}$Ref.\,\onlinecite{ga09} and $^{e}$Ref.\,\onlinecite{sza13}.
Our calculated hyperfine structure constants agree very well with those determined experimentally in Ref.~\onlinecite{fel09}.
\label{tab_C13_A_Bulk}}
\end{table}

\section{Summary}\label{sec:summary}
We have studied the influence of extended planar defects on the formation energies, zero-field splittings and hyperfine structures of negatively charged NV centers in diamond crystals. Our study includes the most relevant planar defects in \{111\} orientation in diamond, namely intrinsic stacking faults, extrinsic stacking faults and coherent twin boundaries.
Altogether, there is a tendency of NV$^-$ to be located at ISF, ESF or CTB, since the formation
energy of the NV-center is reduced compared to the bulk crystal for several sites in vicinity of the defect plane. However, for all the considered extended defects the energy differences between axially and basally oriented NV centers are small (0.1 eV or less) so that a preferential alignment of NV centers may not necessarily be expected.

The hyperfine structure constants as well as the zero-field splitting parameters are influenced by the extended defects only in a very short range within only a few atomic double layers.

The axial component $D$ of the ZFS of a NV center, responsible for the singlet-triplet splitting, is at most reduced by about 9\% in the environment of an extended defect. For
basally oriented NV centers the transversal component $E$ of the ZFS is not zero. Depending 
on the site and the extended defect type we obtain $E$-values of  
about 250$\pm$30 MHz which cause 
a splitting of the triplet levels. The key structural elements causing the changes and the splittings
are local hexagonal ABA stacking sequences, like in hexagonal Lonsdaleite, which disturb the ABC stacking sequence of cubic diamond.

For axially oriented NV$^-$ centers the HFS constants $A_{ii}$ for $^{14}$N increase or decrease by approximately 10\% and for basal orientation the $A_{ii}$ almost vanish at local hexagonally stacked layers. 
The HFS constants $A_{ii}$ for $^{13}$C decrease by about 20 MHz for axially oriented NV$^-$ centers near the extended defects and for basal orientation the $A_{ii}$ decrease or increase (ranging from 101 Mhz to 131 MHz for $A_{11}$ and $A_{22}$ and from 185 MHz to 212 MHz for $A_{33}$) depending on the reduction or accumulation of charge at the site of the carbon probe atom. 
All of the changes of the hyperfine constants are related to the appearance of the key structural element, namely local hexagonal ABA stacks.

In summary, we believe that our data for the ZFS and the HFS helps to interpret spectra of NV doped diamond
samples. On the one hand, variations in the measured spectra of individual NV-centers may be traced back to the presence of extended defects. On the other hand, for sufficiently high densities of NV centers the spectral data may allow to draw conclusions about the density of extended defects with the NV centers acting as probes in the sample. In the context of using NV centers as qubits, our results may help to clarify the interaction of the S= 1 electron spin of the ground state with neighboring $^{13}$C isotopes possessing a nuclear spin $I=$ 1/2.

\section{Acknowledgments}

Financial support for this work was provided by the Fraunhofer LIGHTHOUSE PROJECT QMag.

\end{document}